\documentclass[11pt]{article}

\usepackage[utf8]{inputenc}
\usepackage[T1]{fontenc}
\usepackage{amsmath,amssymb,amsfonts}
\usepackage{graphicx}
\usepackage{booktabs}
\usepackage{hyperref}
\usepackage{cleveref}
\usepackage[margin=1in]{geometry}
\usepackage{listings}
\usepackage{xcolor}
\usepackage{natbib}
\usepackage{tikz}
\usetikzlibrary{arrows.meta, positioning, calc}
\crefname{lstlisting}{Listing}{Listings}
\Crefname{lstlisting}{Listing}{Listings}
\crefname{equation}{Eqn.}{Eqns.}
\Crefname{equation}{Eqn.}{Eqns.}

\definecolor{codebg}{RGB}{245,245,245}
\definecolor{codegreen}{RGB}{0,128,0}
\definecolor{codepurple}{RGB}{128,0,128}
\title{\texttt{BSTModelKit.jl}: A Julia Package for Constructing, Solving, and Analyzing Biochemical Systems Theory Models}
\author{
    Sandra Vadhin\thanks{\href{mailto:sv442@cornell.edu}{sv442@cornell.edu}}\hspace{2pt} and
    Jeffrey D. Varner\thanks{Corresponding author: \href{mailto:jdv27@cornell.edu}{jdv27@cornell.edu}} \\
    \textit{Robert Frederick Smith School of Chemical and Biomolecular Engineering} \\
    \textit{Cornell University, Ithaca, NY 14853, USA}
}
\date{}

\begin{document}
\maketitle

% ============================================================================ %
\begin{abstract}
We present \texttt{BSTModelKit.jl}, an open-source Julia package for constructing, solving, and analyzing Biochemical Systems Theory (BST) models of biochemical networks. The package implements S-system representations, a canonical power-law formalism for modeling metabolic and regulatory networks. \texttt{BSTModelKit.jl} provides a declarative model specification format, dynamic simulation via ordinary differential equation (ODE) integration, steady-state computation, and global sensitivity analysis using the Morris and Sobol methods. The package leverages the Julia scientific computing ecosystem, in particular the SciML suite of differential equation solvers, to provide efficient and flexible model analysis tools. We describe the mathematical formulation, software design, and demonstrate the package capabilities with illustrative examples.
\end{abstract}


% ============================================================================ %
\section{Introduction}\label{sec:introduction}
Mathematical modeling of biochemical networks is essential for understanding the complex dynamics of living systems. Among the many formalisms available, Biochemical Systems Theory (BST) occupies a distinctive position: it provides a canonical mathematical structure grounded in a power-law approximation of enzyme kinetics that is both analytically tractable and broadly applicable. BST was introduced by Savageau in a series of foundational papers \citep{Savageau1969a,Savageau1969b,Savageau1970} and later elaborated in a comprehensive monograph \citep{Savageau1976}. The key insight underlying BST is that, in the vicinity of an operating point, the net rate of production or consumption of any biochemical species can be well-approximated by a product of power-law functions of the concentrations of the species in the system. This approximation is exact for mass-action kinetics and provides a remarkably good local representation of more complex rate laws, including Michaelis-Menten and Hill-type kinetics \citep{Savageau1976,Voit2000}.

The power-law formalism gives rise to two canonical representations: the Generalized Mass Action (GMA) system, which retains individual reaction terms, and the S-system, which aggregates all production and consumption terms for each species into single power-law expressions \citep{Savageau1987}. The S-system representation has particularly attractive mathematical properties---its steady-state equations can be transformed into a system of linear algebraic equations in logarithmic coordinates, enabling efficient analytical and numerical solution \citep{Savageau1976,Voit2000}---and Savageau and Voit established that BST and metabolic control theory yield equivalent descriptions of metabolic regulation near steady state \citep{Savageau1987}. These properties have led to broad application across biological domains. In metabolic engineering, S-system models have been used to characterize and optimize pathways such as glycolysis in \textit{Saccharomyces cerevisiae} \citep{Cascante1995} and to develop systematic strategies for representing reversible metabolic pathways \citep{Sorribas1989}. Savageau and Alves applied ensemble methods based on S-system models to study systemic properties of metabolic networks \citep{Alves2000}. In gene regulation, BST has provided design principles for understanding the functional logic of elementary gene circuits \citep{Savageau2001}, and in signal transduction, Vera et al.\ demonstrated that S-system models can capture the essential dynamics of signaling cascades with fewer parameters than mechanistic alternatives \citep{Vera2007}. Voit provides comprehensive reviews of both the theory and its applications \citep{Voit2000,Torres2002,Voit2013,Voit2013review}. Despite this rich theoretical foundation and broad applicability, the software landscape for BST modeling has not kept pace. Early computational tools such as PLAS \citep{Voit2000} and the specialized numerical methods of Irvine and Savageau \citep{Irvine1990} were pioneering but are tied to legacy computing environments. General-purpose systems biology platforms such as COPASI \citep{Hoops2006} support a variety of kinetic formalisms but do not provide specialized support for the power-law structure of BST models, nor do they expose the stoichiometric and kinetic order matrices that are central to BST analysis. This gap motivates the development of \texttt{BSTModelKit.jl}, an open-source Julia \citep{Bezanson2017} package that provides a complete workflow for BST modeling: declarative model specification, automated construction of the stoichiometric and kinetic order matrices, dynamic simulation, steady-state computation, and global sensitivity analysis. Julia is well-suited for this purpose, as its multiple dispatch paradigm enables clean abstractions, its just-in-time compilation delivers performance competitive with statically compiled languages, and its scientific computing ecosystem---particularly the SciML differential equation solvers \citep{Rackauckas2017} and \texttt{GlobalSensitivity.jl} \citep{Dixit2022}---provides state-of-the-art numerical methods that \texttt{BSTModelKit.jl} leverages directly.

In this paper, we describe the mathematical foundations of the generalized S-system representation used by \texttt{BSTModelKit.jl} (\Cref{sec:formulation}), the software architecture and design (\Cref{sec:software}), illustrative examples demonstrating the package capabilities (\Cref{sec:examples}), and conclude with a discussion of limitations and future directions (\Cref{sec:discussion}).

% ============================================================================ %
\section{Mathematical Formulation}\label{sec:formulation}
Consider a biochemical network with $n$ dynamic species and $m$ static (externally controlled) species. Let $X_i$ denote the concentration of species $i$. In the S-system formalism, the time evolution of each dynamic species is governed by:
\begin{equation}\label{eq:ssystem}
    \frac{dX_i}{dt} = \alpha_i \prod_{j=1}^{n+m} X_j^{g_{ij}} - \beta_i \prod_{j=1}^{n+m} X_j^{h_{ij}}, \qquad i = 1, \dots, n
\end{equation}
where $\alpha_i$ and $\beta_i$ are non-negative rate constants, $g_{ij}$ and $h_{ij}$ are real-valued kinetic orders, and the products extend over all species, both dynamic and static. The first term represents the aggregate production rate of $X_i$, while the second represents its aggregate consumption rate. The kinetic orders $g_{ij}$ and $h_{ij}$ capture the sensitivity of each flux to changes in species concentrations: positive values indicate activation, negative values indicate inhibition, and zero indicates no dependence.

\texttt{BSTModelKit.jl} uses a generalized matrix representation that decouples the stoichiometry from the kinetics. Rather than lumping all production and consumption into two aggregate terms per species, the system dynamics are expressed as:
\begin{equation}\label{eq:matrix}
    \frac{d\mathbf{x}}{dt} = \mathbf{S} \cdot \mathbf{r}(\mathbf{x}, \mathbf{x}_s) + \mathbf{u}(t)
\end{equation}
where $\mathbf{x} \in \mathbb{R}^n$ is the vector of dynamic species concentrations, $\mathbf{x}_s \in \mathbb{R}^m$ is the vector of static species concentrations, $\mathbf{S} \in \mathbb{R}^{n \times p}$ is the stoichiometric matrix for $p$ reactions, $\mathbf{r} \in \mathbb{R}^p$ is the reaction rate vector, and $\mathbf{u}(t) \in \mathbb{R}^n$ is an optional external input vector. Each reaction rate $r_k$ is computed using the power-law kinetic formalism:
\begin{equation}\label{eq:powerlaw}
    r_k = \alpha_k \prod_{j=1}^{n+m} X_j^{G_{jk}}, \qquad k = 1, \dots, p
\end{equation}
where $\boldsymbol{\alpha} \in \mathbb{R}^p$ is the rate constant vector and $\mathbf{G} \in \mathbb{R}^{(n+m) \times p}$ is the kinetic order (exponent) matrix. This formulation generalizes the classical S-system (\Cref{eq:ssystem}) by allowing arbitrary stoichiometric coefficients and providing an explicit separation between network structure ($\mathbf{S}$) and kinetics ($\mathbf{G}$, $\boldsymbol{\alpha}$). The matrix form also makes the connection to stoichiometric network analysis transparent and facilitates algorithmic construction of models from structured input files.

At steady state, the time derivatives vanish, giving the algebraic condition:
\begin{equation}\label{eq:steadystate}
    \mathbf{0} = \mathbf{S} \cdot \mathbf{r}(\mathbf{x}_{ss}, \mathbf{x}_s) + \mathbf{u}
\end{equation}
\texttt{BSTModelKit.jl} computes steady-state solutions using the \texttt{DynamicSS} algorithm from \texttt{Steady\-State\-Diff\-Eq.jl}, which integrates the ODE system forward in time until the derivatives are sufficiently small. This approach is robust for systems where direct algebraic solution of \Cref{eq:steadystate} is intractable, as is typical for power-law systems with many interacting species.

To quantify the influence of model parameters on system behavior, \texttt{BSTModelKit.jl} provides wrappers for two global sensitivity analysis methods from \texttt{GlobalSensitivity.jl} \citep{Dixit2022}. The Morris method \citep{Morris1991} computes elementary effects by perturbing one parameter at a time along randomized trajectories through the parameter space. For each parameter $\theta_k$, the method estimates the mean $\hat{\mu}_k$ and variance $\hat{\sigma}_k^2$ of the elementary effects; parameters with large $|\hat{\mu}_k|$ have significant influence on the output, while large $\hat{\sigma}_k^2$ indicates nonlinear effects or interactions. The Morris method is computationally inexpensive and is well-suited for screening large parameter spaces to identify the most influential parameters. The Sobol method \citep{Sobol2001} provides a more detailed variance-based decomposition, partitioning the total output variance into contributions from individual parameters (first-order indices $S_i$) and parameter interactions (total-order indices $S_{T_i}$). While more computationally demanding, the Sobol method yields a complete picture of parameter importance and interaction structure.

% ============================================================================ %
\section{Software Design}\label{sec:software}
\texttt{BSTModelKit.jl} is organized around a central \texttt{BSTModel} type that encapsulates all data required to simulate and analyze a BST model, including the lists of dynamic and static species, the stoichiometric matrix $\mathbf{S}$, the kinetic order matrix $\mathbf{G}$, the rate constant vector $\boldsymbol{\alpha}$, initial conditions, and static factor values. The model object is mutable, allowing users to adjust parameters, initial conditions, and matrix entries after construction without rebuilding the model from file.

The package exposes a small public API centered on seven functions; complete documentation including usage examples and function signatures is available at \url{https://varnerlab.org/BSTModelKit.jl/dev/}. The \texttt{build} function constructs a \texttt{BSTModel} from a file, supporting TOML, BST, and JLD2 formats. The \texttt{evaluate} function simulates the dynamic trajectory by integrating the ODE system (\Cref{eq:matrix}) over a user-specified time span, returning vectors of time points and state values. The \texttt{steadystate} function computes the steady-state solution of the system (\Cref{eq:steadystate}). The \texttt{savemodel} and \texttt{loadmodel} functions serialize and deserialize model objects to JLD2 binary files, enabling reproducible workflows. Finally, the \texttt{morris} and \texttt{sobol} functions perform global sensitivity analysis given a user-defined scalar performance function and parameter bounds.

Models are specified in declarative text files that define the species, network connectivity, kinetic dependencies, and stoichiometric coefficients. The recommended format is TOML, which is human-readable and well-supported by Julia's standard library. A TOML model file contains a \texttt{[metadata]} section with author and version information, and a \texttt{[model]} section that lists the dynamic and static species, the reaction connection records (specifying reactants and products for each reaction), the kinetic records (specifying which species appear in the rate law for each reaction), and optional stoichiometric coefficient overrides. An example TOML model file for a linear pathway with feedback inhibition is shown in \Cref{lst:toml}. The package also supports a custom BST text format that uses section markers (e.g., \texttt{\#dynamic::start}/\texttt{\#dynamic::end}) and line comments prefixed with \texttt{//}.

\begin{lstlisting}[caption={TOML model file for a linear pathway with feedback inhibition.},label={lst:toml},language={}]
[metadata]
author = "jdv27@cornell.edu"
version = "0.1"
description = "Linear pathway with feedback"

[model]
list_of_static_species = ["E1", "E2", "E3"]
list_of_dynamic_species = ["X1", "X2", "X3", "X4", "X5"]

list_of_connection_records = [
    "r1::{X1} --> X2",
    "r2::{X2} --> X3",
    "r3::X3 --> X4",
    "r5::X3 --> X5",
    "r0::{} --> X1",
    "r4::X4 --> {}"
]

list_of_kinetics_records = [
    "r0::{}", "r1::{X1,X4,E1}",
    "r2::{X2,E2}", "r3::{X3,E3}",
    "r4::{X4}", "r5::{X3}"
]
\end{lstlisting}

\Cref{tab:api} summarizes the public API.

\begin{table}[ht]
    \centering
    \caption{Public API of \texttt{BSTModelKit.jl}. All functions operate on or produce \texttt{BSTModel} instances.}
    \label{tab:api}
    \small
    \begin{tabular}{lll}
        \toprule
        \textbf{Function} & \textbf{Returns} & \textbf{Description} \\
        \midrule
        \texttt{build(path)} & \texttt{BSTModel} & Construct model from TOML, BST, or JLD2 file \\
        \texttt{evaluate(model; ...)} & $(T, X)$ & Simulate dynamic trajectory via ODE integration \\
        \texttt{steadystate(model; ...)} & $\mathbf{x}_{ss}$ & Compute steady-state concentrations \\
        \texttt{morris(f, L, U; ...)} & $(\hat{\mu}, \hat{\sigma}^2)$ & Morris elementary-effects screening \\
        \texttt{sobol(f, L, U; ...)} & \texttt{SobolResult} & Sobol variance-based sensitivity indices \\
        \texttt{savemodel(path, model)} & --- & Serialize model to JLD2 binary file \\
        \texttt{loadmodel(path)} & \texttt{BSTModel} & Deserialize model from JLD2 binary file \\
        \bottomrule
    \end{tabular}
\end{table}

\texttt{BSTModelKit.jl} builds on several packages from the Julia ecosystem. Dynamic simulation and steady-state computation are handled by \texttt{OrdinaryDiffEq.jl} and \texttt{SteadyStateDiffEq.jl} \citep{Rackauckas2017}, which provide adaptive-step explicit and implicit ODE solvers. Global sensitivity analysis is performed through \texttt{GlobalSensitivity.jl} \citep{Dixit2022}, with quasi-random sampling provided by \texttt{QuasiMonteCarlo.jl}. Model serialization uses \texttt{JLD2.jl}, a Julia-native binary format that preserves type information across save and load cycles.

% ============================================================================ %
\section{Examples}\label{sec:examples}
We demonstrate the capabilities of \texttt{BSTModelKit.jl} with three examples of increasing complexity: dynamic simulation of a feedback-inhibited linear pathway, steady-state analysis of a branched pathway under varying enzyme levels, and global sensitivity analysis using the Morris and Sobol methods. All examples use the TOML model specification format and are available in the package repository.

\subsection{Dynamic simulation of a feedback-inhibited pathway}\label{sec:ex-dynamic}
We considered a five-species linear pathway in which a precursor $X_1$ was converted through intermediates $X_2$ and $X_3$ to a product $X_4$, with a branch producing a byproduct $X_5$ from $X_3$. Three static enzymes $E_1$, $E_2$, and $E_3$ catalyzed reactions $r_1$, $r_2$, and $r_3$, respectively. Product $X_4$ exerted feedback inhibition on $r_1$ by appearing in its rate law with a negative kinetic order ($g_{X_4,r_1} = -0.5$), so that as $X_4$ accumulated, the flux from $X_1$ to $X_2$ was progressively suppressed. The model was specified in the TOML format shown in \Cref{lst:toml}, with static factor values $E_1 = E_2 = E_3 = 1.0$, rate constants $\boldsymbol{\alpha} = (10, 10, 10, 0.1, 0, 3)$ for reactions $(r_1, r_2, r_3, r_5, r_0, r_4)$, and initial conditions near zero for all species. Rather than using a constant source rate, we drove the production of $X_1$ through a time-varying external input function $u_1(t)$ that applied two square pulses of different amplitude: a high pulse ($u_1 = 10$) from $t = 5$ to $t = 15$, followed by a return to baseline ($u_1 = 1$), and then a medium pulse ($u_1 = 5$) from $t = 25$ to $t = 35$. This input protocol exercised the feedback mechanism under two different loading conditions.

\begin{figure}[ht]
    \centering
    \begin{tikzpicture}[
        species/.style={circle, draw, thick, minimum size=8mm, font=\footnotesize},
        enzyme/.style={font=\scriptsize\itshape, text=gray},
        arr/.style={-{Stealth[length=2.5mm]}, thick},
        inh/.style={-{Tee Barb[length=2.5mm, width=3mm]}, thick, red!70!black, dashed},
        node distance=14mm
    ]
        \node[species] (X1) {$X_1$};
        \node[species, right=of X1] (X2) {$X_2$};
        \node[species, right=of X2] (X3) {$X_3$};
        \node[species, below right=10mm and 14mm of X3] (X4) {$X_4$};
        \node[species, above right=10mm and 14mm of X3] (X5) {$X_5$};

        \node[left=8mm of X1] (src) {};
        \node[right=8mm of X4] (sink) {};

        \draw[arr] (src) -- node[above, font=\scriptsize] {$r_0$} (X1);
        \draw[arr] (X1) -- node[above, font=\scriptsize] {$r_1$} (X2);
        \draw[arr] (X2) -- node[above, font=\scriptsize] {$r_2$} (X3);
        \draw[arr] (X3) -- node[below left, font=\scriptsize] {$r_3$} (X4);
        \draw[arr] (X3) -- node[above left, font=\scriptsize] {$r_5$} (X5);
        \draw[arr] (X4) -- node[above, font=\scriptsize] {$r_4$} (sink);

        \node[enzyme, above=2mm of {$(X1)!0.5!(X2)$}] {$E_1$};
        \node[enzyme, above=2mm of {$(X2)!0.5!(X3)$}] {$E_2$};
        \node[enzyme, right=1mm of {$(X3)!0.5!(X4)$}, font=\scriptsize\itshape, text=gray] {$E_3$};

        \draw[inh] (X4) .. controls +(-0.3,-1.0) and +(0,-1.2) .. ($(X1)!0.5!(X2)+(0,-0.3)$);
    \end{tikzpicture}

    \vspace{2mm}

    \includegraphics[width=0.85\textwidth]{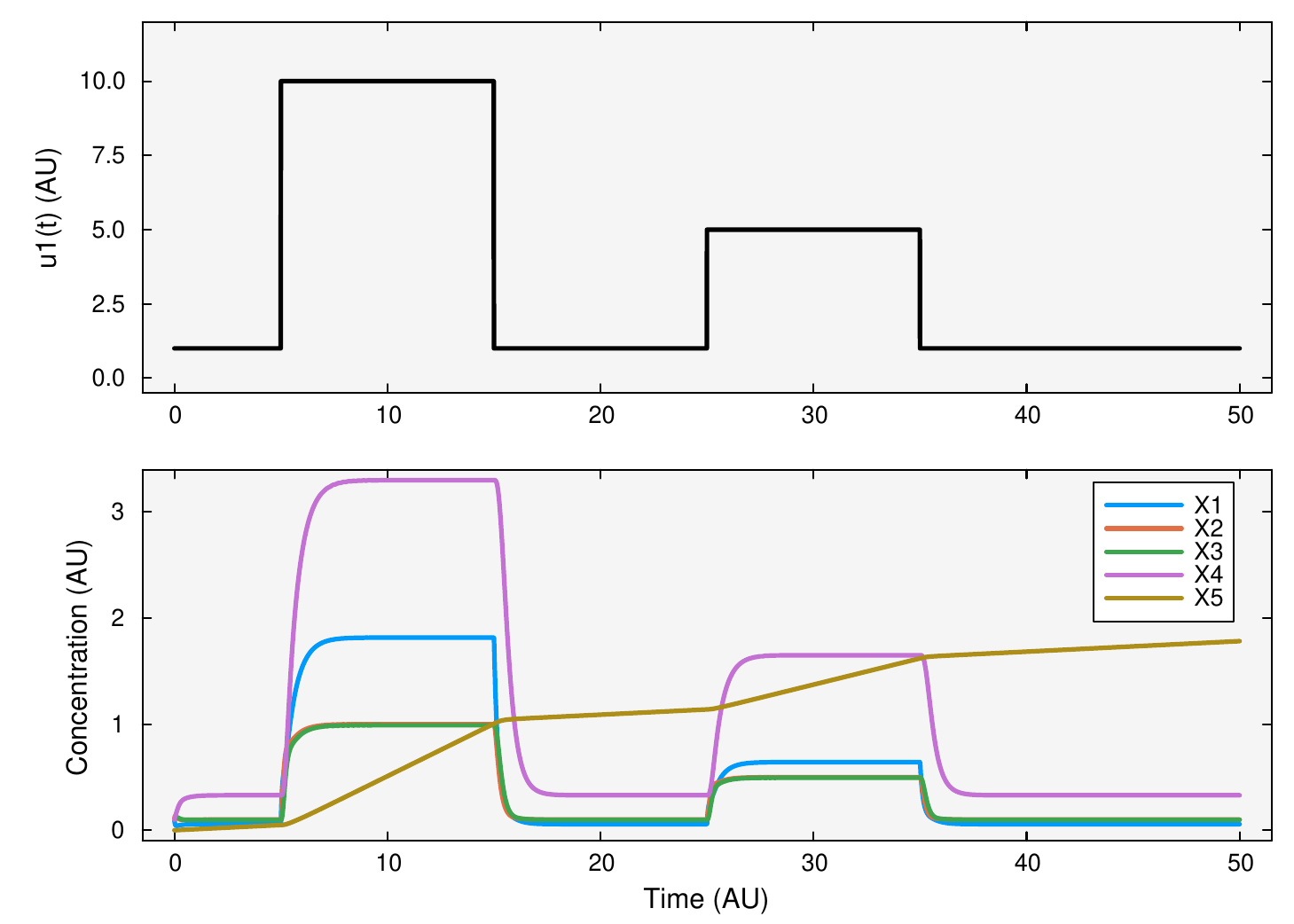}
    \caption{Dynamic simulation of a feedback-inhibited linear pathway driven by a pulsed input. Top panel: reaction network schematic showing a linear pathway from $X_1$ to $X_4$ with a branch to byproduct $X_5$; solid arrows denote mass flow catalyzed by enzymes $E_1$, $E_2$, and $E_3$, while the dashed red line indicates feedback inhibition of $r_1$ by the product $X_4$ ($g_{X_4,r_1} = -0.5$). Middle panel: the time-varying input function $u_1(t)$ applied to the production of $X_1$, consisting of a high pulse ($u_1 = 10$, $t = 5$--$15$) and a medium pulse ($u_1 = 5$, $t = 25$--$35$) separated by returns to baseline ($u_1 = 1$). Bottom panel: simulated concentration trajectories for all five species over $t \in [0, 50]$; $X_4$ accumulation during each pulse suppresses $r_1$ via feedback, limiting upstream buildup, while the byproduct $X_5$ accumulates monotonically because it lacks a degradation pathway.}
    \label{fig:feedback}
\end{figure}

We simulated the dynamic trajectory using the \texttt{evaluate} function and observed that the feedback loop attenuated the response to both pulses (\Cref{fig:feedback}). During the high pulse, $X_1$ rose rapidly, the intermediates $X_2$ and $X_3$ reached approximately unit concentration, and $X_4$ accumulated to its highest level ($\approx 3.3$), which in turn suppressed $r_1$ and limited further buildup of $X_1$. During the medium pulse, the system responded proportionally---the $X_1$ and $X_4$ peaks were roughly half their previous values---demonstrating that the feedback mechanism operated across a range of input magnitudes. The byproduct $X_5$, which had no degradation pathway, accumulated monotonically and ratcheted upward with each pulse, a behavior that would be important to consider in a metabolic engineering context.

\subsection{Steady-state analysis of a branched pathway}\label{sec:ex-steadystate}
We next considered a branched pathway in which a source produced species $A$, which was converted to $B$, then to $C$, at which point the pathway branched: $C$ was converted to $D$ via reaction $r_3$ (catalyzed by enzyme $E_3$) or to $E$ via reaction $r_4$ (catalyzed by enzyme $E_4$). Both $D$ and $E$ were degraded by first-order reactions $r_5$ and $r_6$. Two feedback loops were present: $E$ inhibited $r_1$ ($g_{E,r_1} = -0.5$) and $D$ inhibited $r_2$ ($g_{D,r_2} = -0.5$). This network topology is common in amino acid and nucleotide biosynthesis, where branch point enzymes control the distribution of flux between competing product pathways. The TOML specification for this network is shown in \Cref{lst:toml-branched}; the branch at $C$ is expressed naturally as two separate connection records (\texttt{r3::C --> D} and \texttt{r4::C --> E}), and the feedback dependencies appear in the kinetics records for $r_1$ and $r_2$. To illustrate how enzyme levels at the branch point controlled metabolic flux distribution, we swept the concentration of $E_3$ from 0.1 to 5.0 while holding $E_4 = 1.0$ fixed and computed the steady-state concentrations using the \texttt{steadystate} function.

\begin{lstlisting}[caption={TOML model file for a branched pathway with dual feedback inhibition.},label={lst:toml-branched},language={},float=ht]
[metadata]
author = "jdv27@cornell.edu"
date = "12/17/22"
version = "0.1"
description = "Branched pathway with feedback"

[model]
list_of_static_species = ["E1", "E2", "E3", "E4"]
list_of_dynamic_species = ["A", "B", "C", "D", "E"]

list_of_connection_records = [
    "r1::A --> B",
    "r2::B --> C",
    "r3::C --> D",
    "r4::C --> E",
    "r0::{} --> A",
    "r5::D --> {}",
    "r6::E --> {}"
]

list_of_kinetics_records = [
    "r0::{}",
    "r1::{A,E,E1}",
    "r2::{B,D,E2}",
    "r3::{C,E3}",
    "r4::{C,E4}",
    "r5::{D}",
    "r6::{E}"
]
\end{lstlisting}

\begin{figure}[ht]
    \centering
    \begin{tikzpicture}[
        species/.style={circle, draw, thick, minimum size=8mm, font=\footnotesize},
        enzyme/.style={font=\scriptsize\itshape, text=gray},
        arr/.style={-{Stealth[length=2.5mm]}, thick},
        inh/.style={-{Tee Barb[length=2.5mm, width=3mm]}, thick, red!70!black, dashed},
        node distance=14mm
    ]
        \node[species] (A) {$A$};
        \node[species, right=of A] (B) {$B$};
        \node[species, right=of B] (C) {$C$};
        \node[species, above right=8mm and 14mm of C] (D) {$D$};
        \node[species, below right=8mm and 14mm of C] (E) {$E$};

        \node[left=8mm of A] (src) {};
        \node[right=8mm of D] (sinkD) {};
        \node[right=8mm of E] (sinkE) {};

        \draw[arr] (src) -- node[above, font=\scriptsize] {$r_0$} (A);
        \draw[arr] (A) -- node[above, font=\scriptsize] {$r_1$} (B);
        \draw[arr] (B) -- node[above, font=\scriptsize] {$r_2$} (C);
        \draw[arr] (C) -- node[above left=-1mm, font=\scriptsize] {$r_3$} (D);
        \draw[arr] (C) -- node[below left=-1mm, font=\scriptsize] {$r_4$} (E);
        \draw[arr] (D) -- node[above, font=\scriptsize] {$r_5$} (sinkD);
        \draw[arr] (E) -- node[above, font=\scriptsize] {$r_6$} (sinkE);

        \node[enzyme, above=2mm of {$(C)!0.5!(D)$}] {$E_3$};
        \node[enzyme, below=2mm of {$(C)!0.5!(E)$}] {$E_4$};

        \draw[inh] (E) .. controls +(-1.5,-0.6) and +(-0.3,-1.0) .. ($(A)!0.5!(B)+(0,-0.3)$);
        \draw[inh] (D) .. controls +(-1.5,0.6) and +(-0.3,1.0) .. ($(B)!0.5!(C)+(0,0.3)$);
    \end{tikzpicture}

    \vspace{2mm}

    \includegraphics[width=0.85\textwidth]{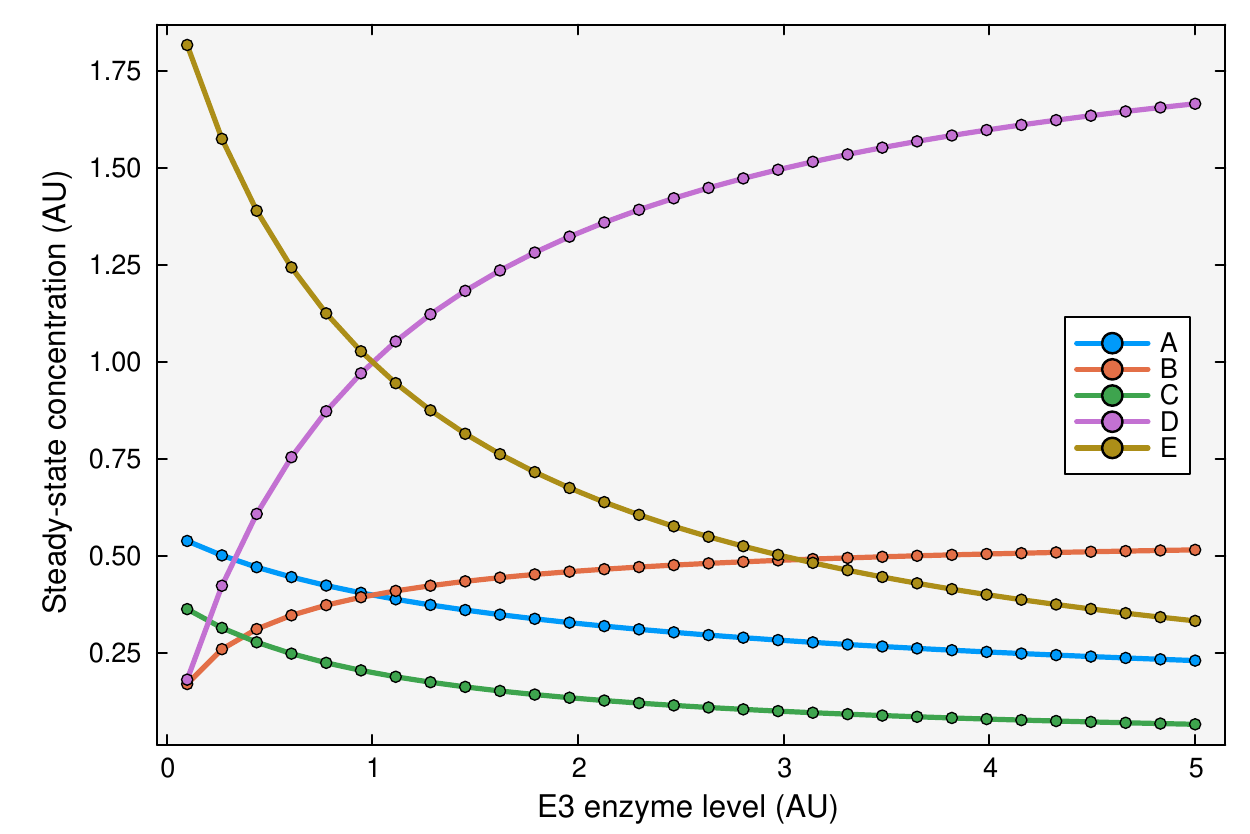}
    \caption{Steady-state analysis of a branched pathway under varying enzyme levels. Top panel: reaction network schematic showing a linear pathway from source to $C$, where the pathway branches to $D$ (via $r_3$, catalyzed by $E_3$) and $E$ (via $r_4$, catalyzed by $E_4$); both $D$ and $E$ are degraded by first-order reactions $r_5$ and $r_6$, respectively; dashed red lines indicate feedback inhibition of $r_1$ by $E$ ($g_{E,r_1} = -0.5$) and $r_2$ by $D$ ($g_{D,r_2} = -0.5$). Bottom panel: steady-state concentrations of all five species as a function of the enzyme level $E_3$ (swept from 0.1 to 5.0, with $E_4 = 1.0$ held fixed), computed using the \texttt{steadystate} function. At low $E_3$, flux is directed predominantly toward $E$; as $E_3$ increases, the $D$ and $E$ curves cross near $E_3 \approx 1$ and the flux redistributes to favor the $D$ branch. The upstream species $A$ and $B$ decrease at high $E_3$ due to modulation by the feedback loops.}
    \label{fig:steadystate}
\end{figure}

We found that increasing $E_3$ systematically redirected flux from the $E$ branch to the $D$ branch (\Cref{fig:steadystate}). At low $E_3$, reaction $r_3$ was slow relative to $r_4$, so most flux was directed toward $E$; the steady-state concentration of $E$ exceeded 1.8 while $D$ was below 0.2. As $E_3$ increased, the flux redistributed: the $D$ and $E$ curves crossed near $E_3 \approx 1$, and at high $E_3$ the situation reversed, with $D$ reaching approximately 1.7 and $E$ dropping below 0.4. The branch point metabolite $C$ decreased monotonically with increasing $E_3$, as it was consumed more rapidly. The upstream species $A$ and $B$ also adjusted through the feedback loops, decreasing as the changing $D$ and $E$ levels modulated $r_1$ and $r_2$. This example demonstrated how \texttt{BSTModelKit.jl} could be used to explore the relationship between enzyme levels and steady-state flux distribution, a central concern in metabolic engineering.

\subsection{Global sensitivity analysis}\label{sec:ex-sensitivity}
Finally, we performed a global sensitivity analysis to identify which parameters most influenced the time-integrated concentration of $X_4$ in the feedback-inhibited pathway from \Cref{sec:ex-dynamic}. The seven parameters varied were the six rate constants $\alpha_{r_0}, \alpha_{r_1}, \ldots, \alpha_{r_5}$ (with bounds at $\pm 50\%$ of their nominal values) and the feedback kinetic order $g_{X_4,r_1}$ (varied from $-2$ to $0$). The scalar performance metric was $J = \int_0^{20} X_4(t)\,dt$, representing the cumulative exposure to the product. We applied both the Morris screening method (500 trajectories) and the Sobol variance decomposition (1000 quasi-random samples).

\begin{figure}[ht]
    \centering
    \includegraphics[width=\textwidth]{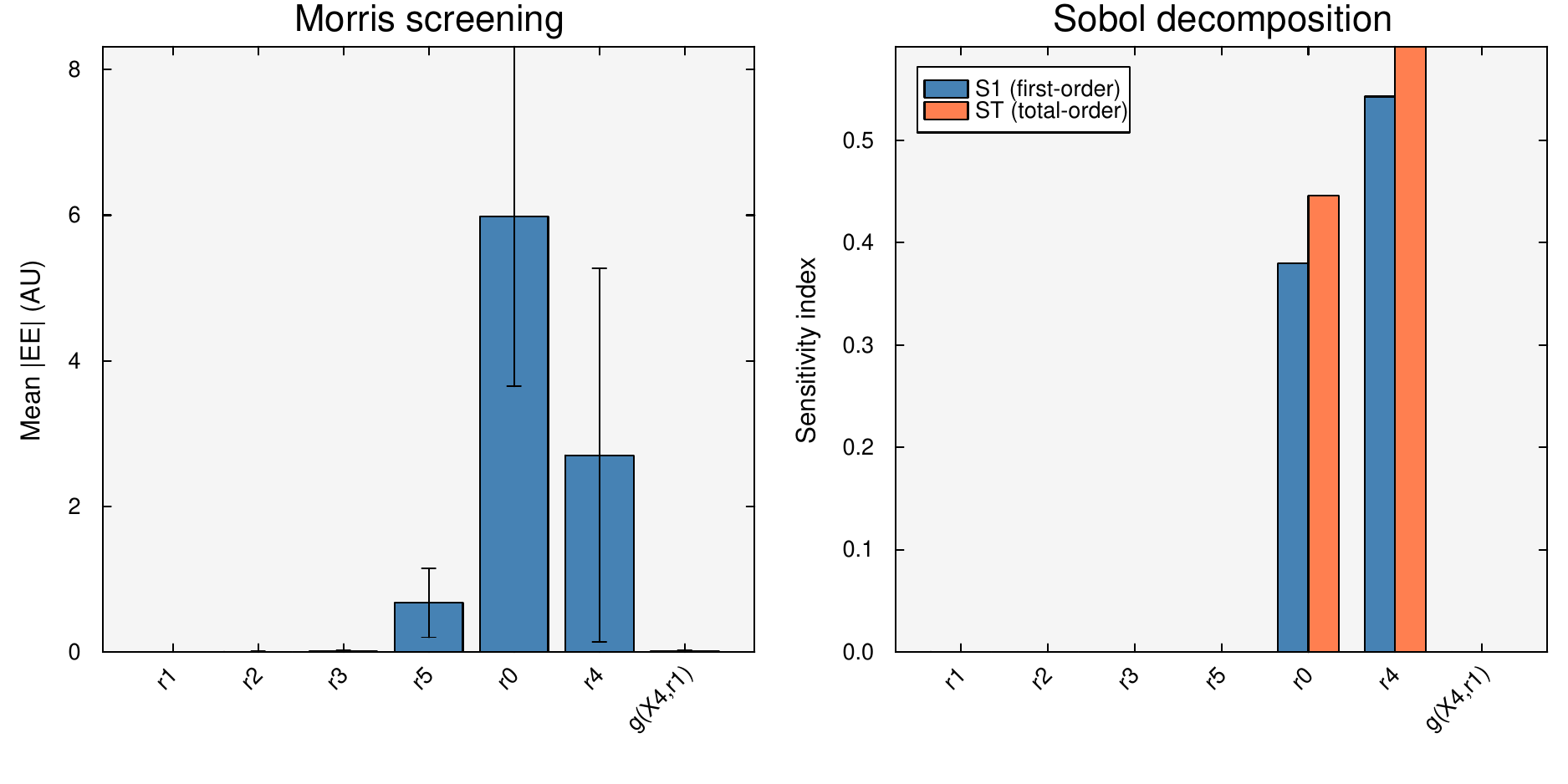}
    \caption{Global sensitivity analysis of the time-integrated $X_4$ concentration ($J = \int_0^{20} X_4\,dt$) for the feedback-inhibited pathway shown in \Cref{fig:feedback}. Seven parameters were varied: six rate constants ($\alpha_{r_1}, \alpha_{r_2}, \alpha_{r_3}, \alpha_{r_5}, \alpha_{r_0}, \alpha_{r_4}$; bounds at $\pm 50\%$ of nominal) and the feedback kinetic order $g_{X_4,r_1}$ (varied from $-2$ to $0$). Left panel: Morris screening (500 trajectories) showing mean absolute elementary effects ($\mu^{*}$) with standard deviation error bars for each parameter; $\alpha_{r_0}$ and $\alpha_{r_4}$ dominate. Right panel: Sobol variance decomposition (1000 quasi-random samples) showing first-order ($S_1$) and total-order ($S_T$) sensitivity indices with confidence intervals; $\alpha_{r_4}$ ($S_1 \approx 0.54$, $S_T \approx 0.59$) and $\alpha_{r_0}$ ($S_1 \approx 0.38$, $S_T \approx 0.45$) account for the majority of output variance, while the gap between $S_1$ and $S_T$ indicates mild parameter interactions.}
    \label{fig:sensitivity}
\end{figure}

We found that the two methods yielded consistent conclusions, identifying the source and degradation rates as the dominant parameters (\Cref{fig:sensitivity}). The Morris elementary effects identified $\alpha_{r_0}$ (source rate) and $\alpha_{r_4}$ (product degradation rate) as the most influential parameters, with $\alpha_{r_5}$ (branch rate) a distant third; the upstream reaction rate constants and the feedback kinetic order had negligible influence. The Sobol analysis confirmed this ranking: $\alpha_{r_4}$ had the largest first-order index ($S_1 \approx 0.54$) and total-order index ($S_T \approx 0.59$), followed by $\alpha_{r_0}$ ($S_1 \approx 0.38$, $S_T \approx 0.45$). The gap between $S_1$ and $S_T$ for both parameters indicated mild interactions, consistent with the multiplicative structure of the power-law kinetics. These results are intuitive: the integrated $X_4$ concentration was controlled primarily by how fast $X_4$ was produced (governed by the source feeding the pathway) and how fast it was removed (governed by $r_4$), while the intermediate conversion steps operated fast enough that their rate constants did not limit the overall accumulation.

% ============================================================================ %
\section{Discussion}\label{sec:discussion}
We have presented \texttt{BSTModelKit.jl}, an open-source Julia package that provides a complete workflow for Biochemical Systems Theory modeling: declarative model specification in TOML or custom text formats, automated construction of the stoichiometric and kinetic order matrices, dynamic simulation via ODE integration, steady-state computation, and global sensitivity analysis using the Morris and Sobol methods. The package leverages the Julia scientific computing ecosystem---particularly \texttt{OrdinaryDiffEq.jl}, \texttt{SteadyStateDiffEq.jl} \citep{Rackauckas2017}, and \texttt{GlobalSensitivity.jl} \citep{Dixit2022}---to provide efficient and numerically robust solvers while maintaining a small, focused API. The three examples presented in \Cref{sec:examples} illustrate the range of analyses that can be performed: dynamic simulation of feedback-regulated pathways under time-varying inputs, systematic exploration of how enzyme levels at metabolic branch points control steady-state flux distribution, and identification of the most influential parameters through global sensitivity analysis.

It is worth noting what \texttt{BSTModelKit.jl} deliberately does not attempt. The package is not a general-purpose systems biology platform: it does not provide a graphical user interface, SBML import/export, or stochastic simulation. It does not perform parameter estimation, model selection, or Bayesian inference. These are mature capabilities available in tools such as COPASI \citep{Hoops2006} and the broader Julia SciML ecosystem. Instead, \texttt{BSTModelKit.jl} occupies a focused niche---programmatic construction, simulation, and sensitivity analysis of S-system models---and is designed to be composed with other Julia packages rather than to replace them.

Several limitations of the current implementation suggest directions for future development. First, \texttt{BSTModelKit.jl} currently supports only the S-system representation; extending the package to support the Generalized Mass Action (GMA) formalism would enable modeling of systems where aggregation of production and consumption terms into single power-law expressions is not appropriate, such as networks with multiple independent production pathways for a single species. Second, the package does not currently provide local sensitivity analysis; integration with Julia's automatic differentiation ecosystem (e.g., \texttt{ForwardDiff.jl}) would enable efficient computation of sensitivity coefficients and logarithmic gains, quantities that are central to the classical BST analysis framework \citep{Savageau1976,Voit2000}. Third, parameter estimation from experimental data is not yet supported; combining the power-law model structure with gradient-based optimization or Bayesian inference methods available in the Julia ecosystem would make it possible to calibrate BST models directly from time-series measurements. Finally, integration with \texttt{ModelingToolkit.jl} could enable symbolic simplification, automatic Jacobian generation, and compilation of optimized model code, further improving performance for large-scale systems. We anticipate that these extensions, together with the growing Julia ecosystem for scientific computing, will make \texttt{BSTModelKit.jl} a useful tool for researchers applying BST to problems in systems biology and metabolic engineering.

% ============================================================================ %
\section*{Acknowledgments}
This work was supported by the National Institutes of Health (NIH) National Heart, Lung, and Blood Institute (NHLBI) under grants R33 HL141787 (The Interaction of Basal Risk, Pharmacological Ovulation Induction, Pregnancy and Delivery on Hemostatic Balance; PIs I.\ Bernstein and T.\ Orfeo) and R01 HL71944 (The Pregnancy Phenotype and Predisposition to Preeclampsia; PI I.\ Bernstein).

% ============================================================================ %
\section*{Data and Code Availability}
\texttt{BSTModelKit.jl} is freely available under the MIT license at \url{https://github.com/varnerlab/BSTModelKit.jl}. The package can be installed from the Julia REPL by entering package mode (\texttt{]} key) and running:
\begin{lstlisting}[language={},numbers=none,backgroundcolor=\color{codebg}]
add BSTModelKit
\end{lstlisting}
\noindent Alternatively, the development version can be installed directly from GitHub:
\begin{lstlisting}[language={},numbers=none,backgroundcolor=\color{codebg}]
add https://github.com/varnerlab/BSTModelKit.jl.git
\end{lstlisting}
\noindent Documentation is available at \url{https://varnerlab.org/BSTModelKit.jl/dev/}. All example code and model files used in this paper are included in the \texttt{paper/code} directory of the repository.

% ============================================================================ %
\bibliographystyle{plainnat}
\bibliography{references}

\end{document}